\documentstyle[11pt,cs11,html,epsf]{article}

\begin{document}

\title{Chromospherically active binaries
members of young stellar kinematic groups}

\author{D. Montes, 
M.J. Fern\'andez-Figueroa,
E. De Castro, M.~Cornide, A.~Latorre}
\affil{Departamento de Astrof\'{\i}sica, 
Univ. Complutense de Madrid, Spain}

% Notice that some of these authors have alternate affiliations, which
% are identified by the \altaffilmark after each name.  The actual alternate
% affiliation information is typeset in footnotes at the bottom of the
% first page, and the text itself is specified in \altaffiltext commands.
% There is a separate \altaffiltext for each alternate affiliation
% indicated above.

% The nice thing about this method is that it saves space on the first page!

%\altaffiltext{1}{Guest observer at McDonald Observatory}

% BUT if you've used \altaffiltext and you think you'll be adding 
% *footnotes*, then you need to update the footnote counter!

%\setcounter{footnote}{3}

% The abstract is entered in a LaTeX "environment", designated with paired
% \begin{abstract} -- \end{abstract} commands.  Other environments are
% identified by the name in the curly braces.

\begin{abstract}

We present a kinematic study of a large sample of 
chromospherically active binaries (CAB)
in order to determine their membership to representative
young disk stellar kinematic groups: the
Local Association (Pleiades moving group, 20 - 150 Myr),
Ursa Mayor group (Sirius supercluster, 300 Myr),
 Hyades supercluster (800 Myr),
IC 2391 supercluster (35 Myr) and
Castor moving group (200 Myr).
Precise measurements of proper motions and parallaxes taken from
Hipparcos Catalogue and published radial velocity measurements
are used to calculate Galactic space motions (U, V, W)
in order to determine the membership of the selected stars to the different
stellar kinematic groups.

\end{abstract}

% Keywords should be included, but they are not printed in the hardcopy.
% They will be used by the Editors to help organize poster papers by
% category though!


% That's it for the front matter.  On to the main body of the paper.
% We'll only put in tutorial remarks at the beginning of each section
% so you can see entire sections together.

% 
% OK - to make things easier for the Editors, we're going to put
% all of our object aliases up front since we only have to declare
% them once in the paper.  Some people prefer to use NGC 7078 for
% M 15, but we like good old Messier, so that's what we'll index by.
% But we'll cross-reference it here so that people who do like NGC 7078
% won't have to remember that it's also M 15!
%
% Remember - we identify objects by putting an asterisk in front of the name!
%
%\index{*BF Lyn|HD 80715}

\section{Introduction}

Activity-rotation and activity-age relationships have been found in many
studies of late-type stars.
The rotation rate moderates the dynamo mechanism which generates and amplifies
the magnetic fields in the convective zone, but there is a further relationship
between rotation and age.
Rotation rates decline with age because stars lose angular momentum through the coupling of the magnetic field and stellar mass loss, and thus there is an 
indirect trend of decreasing activity with increasing age.  
Chromospherically active binaries (CAB) are
detached binary systems with cool components
characterized by strong chromospheric, transition region, and coronal activity.
CAB can lose angular momentum, but maintain high rotation rates and activity 
levels by a decrease in their component separation (synchronization of rotation and orbital periods). Samples of CAB with the same age are of maximum interest
to better understand the magnetic activity of these systems.

Some late-type spectroscopic binaries have been identified as members of well 
known open clusters (Montes 1999, and references therein), but only a few are 
well known CAB.
Stellar kinematic groups (SKG) are kinematically coherent groups of stars that 
share a common origin, and thus offer another way to compile samples of stars 
with the same age. The youngest and best documented SKG are: 
the Hyades supercluster (Eggen 1992b) associated with 
the Hyades cluster (600 Myr),
the Ursa Mayor group (Sirius supercluster) (Eggen 1992a, 1998;
 Soderblom \& Mayor 1993) associated with the UMa cluster
(300 Myr), the Local Association or Pleiades moving group associated with 
the Pleiades and several other young open clusters and associations
(age ranges from about 20 to 150 Myr) (Eggen 1992c), 
the IC 2391 supercluster (35-55 Myr) (Eggen 1995), and
the Castor moving group (200 Myr) (Barrado y Navascu\'es 1998).  
The existence of these SKG  has been rather controversial in the literature,
but recent studies (Chereul et al. 1999, Dehnen 1998, Asiain et al. 1999, 
Skuljan et al. 1999) using astrometric data taken from Hipparcos
not only confirm the existence of classical young moving groups, but also 
detect finer structures in space velocity and age.
Well known members of these SKG are mainly early-type stars and few studies 
have been centered in late-type stars
(see Montes et al. 1999, and Montes 2000 (this proceedings)).  
In this contribution we present a kinematic study of a large sample of CAB 
in order to determine their membership to representative
young disk SKG.  Precise measurements of proper motions and parallaxes taken 
from Hipparcos Catalogue and published radial velocity measurements are used 
to calculate Galactic space motions (U, V, W).

%----------------------------------
\begin{figure}
\plotone{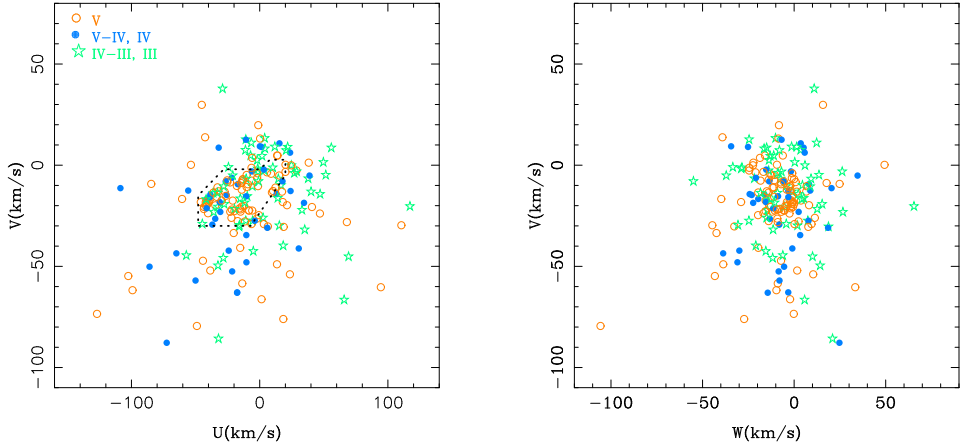}
\caption{The (U, V) and (W, V) planes (Boettlinger Diagram)
for the whole sample.
We have divided the sample in three groups according with their
luminosity class (V, IV, and III).
The stars of the three groups are plotted with different symbols and colors}
 \label{fig-1}
\end{figure}
%----------------------------------
%----------------------------------
\begin{figure}
\plotone{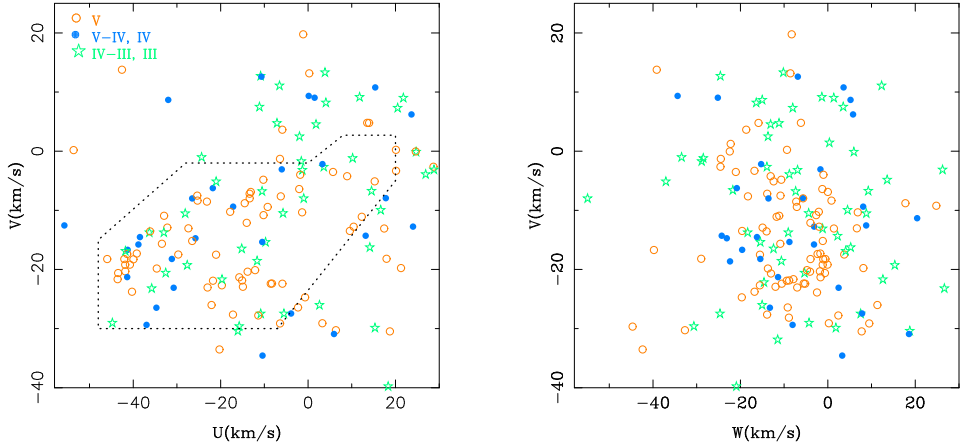}
\caption{Enlargement of the central region of Fig.~\ref{fig-1} 
including the boundaries (dashed line)
that determine the young disk population as defined by Eggen (1984, 1989)}
 \label{fig-2}
\end{figure}
%----------------------------------

%----------------------------------
\section{Sample of CAB and parameters}
%----------------------------------

A total of 205 CAB with complete kinematic input have been
included in this study. The systems  have been selected from
different sources:
\newline
$\bullet$ Previously established members of
stellar kinematic groups based in photometric and kinematic properties
(several papers by Olin Eggen).
\newline
$\bullet$ Possible new candidates found in our previous kinematic study of 
late-type stars (Montes et al. 1999).
\newline
$\bullet$  The 206 CAB included in the "Catalog of Chromospherically Active 
Binary Stars"
(Strassmeier et al. 1993).
\newline
$\bullet$  Some of the CAB included in the candidate list of Strassmeier et al. (1993)
\newline
$\bullet$  Other late-type stars recently identified in the literature
as CAB, including X-ray/EUV selected stars.
(Jeffries et al. 1995, Henry et al. 1995).

In order to determine the membership of this sample to the different stellar 
kinematic groups we have studied the
distribution of stars in the space velocity by calculating the 
 Galactic space-velocity components (U, V , W)
in a right-handed coordinated system (positive in the directions of the 
Galactic center, Galactic rotation, and the
North Galactic Pole, respectively).
The procedures in Johnson \& Soderblom (1987) were used to calculate 
U, V, W, and their associated errors.

Parallaxes and  proper motions are taken from
Hipparcos Catalogue (ESA, 1997);
PPM (Positions and Proper Motions) Catalogue (R\"{o}ser et al, 1994);
ACT Reference Catalog (Urban et al. 1997); and
TCR (Tycho Reference Catalogue) (Hog et al. 1998).
We have only included in the study stars with significant trigonometric 
parallaxes ($\pi$~$\geq$~3$\sigma$$_{\pi}$).
In some cases, when trigonometric parallaxes are not available, we adopted 
spectroscopic parallaxes.
Radial velocities are primarily taken from the system's center of 
mass radial velocity listed
in Strassmeir et al. (1993) catalog or other more recent orbital 
determination found in the literature.
Some radial velocities are also taken from the
WEB (Wilson Evans Batten) compilation (Duflot et al. 1995),
 and from other references given in SIMBAD.

%----------------------------------
\begin{figure}
\plotone{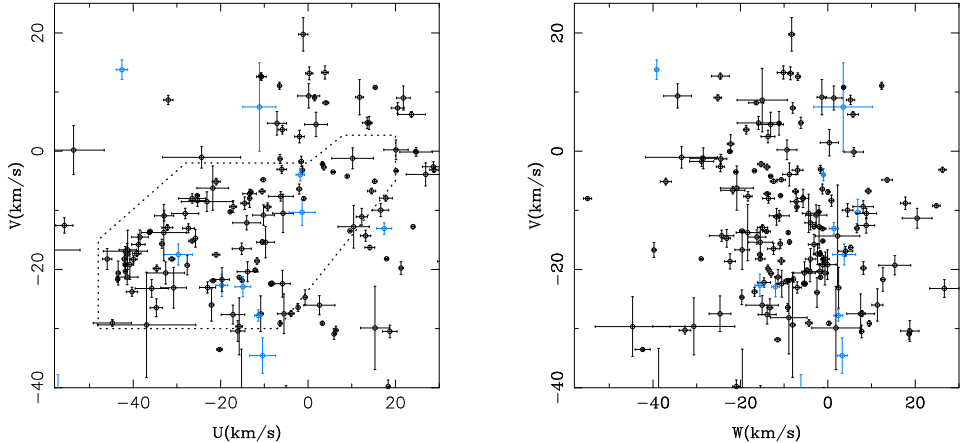}
\caption{Stars with their associated errors in the central
region of the (U, V) and (W, V) planes.
Stars with trigonometric parallaxes have been plotted
in black and stars with spectrocopic parallaxes in blue.}
 \label{fig-3}
\end{figure}
%----------------------------------
%----------------------------------
\section{(U, V) and (W, V) diagrams}
%----------------------------------

The (U, V) and (W, V) planes (Boettlinger Diagram) for the whole sample are 
plotted in  Fig.~\ref{fig-1}.
All the stars fall in the range of U (-130, 120) and V (-90, 40) except
two stars with very large space velocities:
 CM Dra (U = -105.35, V = -119.35) and
 Gl 629.2A (U = -88.24, V = -172.06) which result to be old Population II 
binaries.
We have divided the sample in three groups according to their luminosity 
class (V, IV, and III).
The stars of the three groups are plotted in this figure with different 
simbols and colors.
Fig.~\ref{fig-2} is an enlargement of the central region of Fig.~\ref{fig-1} 
including the boundaries (dashed line)
that determine the young disk population as defined by Eggen (1984, 1989).
As it can be seen in this figure a large number of BY Dra stars 
(luminosity class V) seems to fall inside of the boundaries of the young star 
region, but a considerable number of subgiants and giants also fall in this 
region.  A detailed kinematic study will be the subject of a future work, for 
a previous kinematic study see Eker (1992).

In Fig.~\ref{fig-3} we have plotted each star with its associated error, 
in the central region of the (U, V) and (W, V) diagrams.
Stars with trigonometric parallaxes have been plotted in black and stars with 
spectrocopic parallaxes in blue.
The uncertainties are in general modest, except some cases with large errors, 
which correspond to stars with small trigonometric parallaxes.

\begin{figure}
\plotone{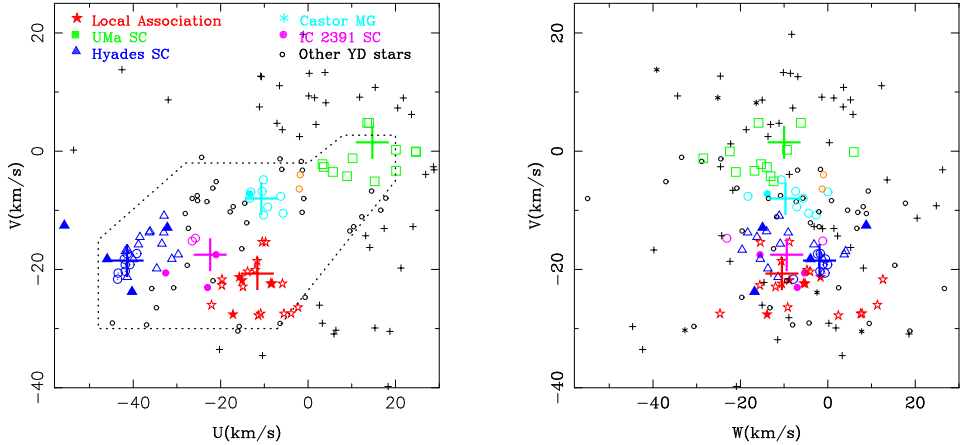}
\caption{(U, V) and (W, V) planes for our star sample.
We plot with different symbols and colors the stars belonging to the different 
SKG, and the other young disk stars.
Filled symbols are member stars (Y in tables) and open symbols are possible 
members (Y? in tables).
Big crosses are plotted in the central position of each group.} 
 \label{fig-4}
\end{figure}

We focus this contribution in the indentification of a preliminary list of 
CAB possible members of some of the five young moving groups above mentioned.
In base of the concentrations in (U, V) and (W, V) planes
around the central position of the different moving groups 
(see Fig.~\ref{fig-4} 
we have classified the stars of our sample as members of one of these moving 
groups or as other possible young disk stars if their classification is not 
clear but it is inside or near the boundaries (dashed line)
of the young disk population.
In Tables 1 to 5
\footnote{Tables 1 to 5 available at 
{\tt http://www.ucm.es/info/Astrof/cabs$\_$yskg.html}}
 we list the candidate stars for each moving group.
We give the name, coordinates (FK5 1950.0), radial velocity (V$_{r}$) and the 
error in km/s, parallax ($\pi$) and the error in milli arc second (mas),
proper motions  $\mu$$_{\alpha}$ and  $\mu$$_{\delta}$ and their errors in 
mas per year (mas/yr), and the U, V, W, calculated components with their 
associated errors in km/s.  In the last column we mark with Y previously 
established members of the stellar kinematic group and Y? possible new 
members in base of their position in the (U, V) plane.

%----------------------------------
%----------------------------------

%----------------------------------
\section{Membership and ages}
%----------------------------------

For some of the CAB listed in Tables 1 to 5,
for which accurate determinations of their stellar parameters are available,
stellar ages have been obtained (Barrado et al. 1994, B94 hereafter) 
by using evolutionary tracks.
In the following we comment some particular cases for each moving group.
\newline
%\vspace{0.1cm}
\underline{\sc  Local Association}
%\vspace{0.2cm}
\newline
Four CABS (V640~Cas~AB, EP~Eri, HD~102077, V772~Her) have been previously 
identified as members of the Local Association.
%\newline
LX Per was  classified as member of $\alpha$ Per open cluster, but the space 
velocities calculated here indicate it is member of the Hyades supercluster.
%\newline
The ages calculated by B94 for TW Lep (94 Myr) and BM CVn (65 Myr) are 
compatible with their membership.
The B94`s age of the doubtful members xi~UMa~B (6 Gyr), 
$\sigma$$^{2}$~CrB (4 Gyr), and ER~Vul (4 Gyr) indicates they are not members.
%\newline
The case of V772 Her is not clear since it seems to be a certain member, 
Batten et al. (1979) suggest an age as the Pleiades,
but the  B94`s age is 3 Gyr.
\newline
%\vspace{0.1cm}
\underline{\sc  IC 2391}
%\vspace{0.2cm}
\newline
Only five CAB could be included in this group of which
TZ~For, HD~54371, HD~58738A are previously established members.
\newline
%\vspace{0.1cm}
\underline{\sc Castor moving group}
%\vspace{0.2cm}
\newline
YY~Gem (Castor C) is one of the stars that define this moving group and
its membership has been confirmed by Barrado y Navascu\'es (1998).
%\newline
VV Mon was initially included as a possible member but the age of 2.6~Gyr 
calculated by B94 indicated it is not a member.
\newline
%\vspace{0.1cm}
\underline{\sc Ursa Mayor group}
%\vspace{0.2cm}
\newline
The age calcuted by B94 for $\epsilon$~UMi (446 Myr) is compatible with 
its membership, however the evolutionary status of
this system is complicated.
%\newline
UV CrB with an age of 5 Gyr (B94) should be rejected as possible member.
\newline
%\vspace{0.1cm}
\underline{\sc Hyades supercluster}
%\vspace{0.2cm}
\newline
Some CAB are previously established members of the Hyades open cluster
(V1136 Tau, V818~Tau, HD~27149, HD~27691, V918~Tau, V808~Tau, QY~Aur) and 
are plotted with a different simbol in Fig.~\ref{fig-4}.
%\newline
Previously established members of the supercluster are:
 ADS~48A (GJ 4A), V471~Tau, and DH~Leo.
%\newline
The age calculated by B94 for 93 Leo (933 Myr) is close to the Hyades ages, 
but the age of HD~131832 (93 Myr) is too young and the ages of 
HD~3196 (1.7 Gyr), RZ Eri (2.2 Gyr), and LU Hya (4.3 Gyr)
are too old to be members.
\newline
%\vspace{0.1cm}
\underline{\sc Other possible young disk stars}
%\vspace{0.2cm}
\newline
In this group of other possible young disk CAB we found several young stars 
as calculated by B94, but also some old stars.

%----------------------------------

% Finally, we have a little acknowledgements section.

\acknowledgments

This work has been supported by the Universidad Complutense de Madrid
and the Spanish Direcci\'{o}n General de  Ense\~{n}anza Superior e
Investigaci\'{o}n Cient\'{\i}fica (DGESIC) under grant PB97-0259.

\end{document}